\documentstyle[epsf]{mn}

\newif\ifAMStwofonts

\def\kms{\relax \ifmmode {\,\rm km\,s}^{-1}\else \,km\,s$^{-1}$\fi}
\def\ha{\relax \ifmmode {\rm H}\alpha\else H$\alpha$\fi}
\def\hb{\relax \ifmmode {\rm H}\beta\else H$\beta$\fi}
\def\hi{\relax \ifmmode {\rm H\,{\sc i}}\else H\,{\sc i}\fi}
\def\hii{\relax \ifmmode {\rm H\,{\sc ii}}\else H\,{\sc ii}\fi}
\def\h2{\relax \ifmmode {\rm H}_2\else H$_2$\fi}
\def\lha{\relax \ifmmode L_{{\rm H}\alpha}\else $L_{{\rm H}\alpha}$\fi}
\def\shi{\relax \ifmmode \sigma_{{\rm HI}}\else $\sigma_{\rm HI}$\fi}
\def\sh2{\relax \ifmmode \sigma_{{\rm H}_2}\else $\sigma_{{\rm H}_2}$\fi}
\def\degr{\hbox{$^\circ$}}
\def\arcmin{\hbox{$^\prime$}}
\def\arcsec{\hbox{$^{\prime\prime}$}}
\def\deg{\hbox{$^\circ$}}
\def\min{\hbox{$^\prime$}}
\def\sec{\hbox{$^{\prime\prime}$}}
\def\fdg{\hbox{$.\!\!^\circ$}}
\def\fs{\hbox{$.\!\!^{\rm s}$}}
\def\farcm{\hbox{$.\mkern-4mu^\prime$}}
\def\farcs{\hbox{$.\!\!^{\prime\prime}$}}
\def\degd#1.#2{ #1\fdg#2 }                 
\def\mind#1.#2{ #1\farcm#2 }               
\def\secd#1.#2{ #1\farcs#2 }               
\def\hhh{\ifmmode {\rm ^h}              
         \else {${\rm ^h}$}
         \fi}
\def\sss{\ifmmode {\rm ^s}              
         \else {${\rm ^s}$}
         \fi}
\def\hms#1h#2m#3s{                      
                  \relax
                  \ifmmode #1^{\rm h}\,#2^{\rm m}\,#3^{\rm s}
                  \else \hbox{$#1^{\rm h}\,#2^{\rm m}\,#3^{\rm s}$}
                  \fi
                 }
\def\dms#1d#2m#3s{                      
                  \relax
                  #1\degr\,#2\arcmin\,#3\arcsec 
                 }
\def\hmsd#1h#2m#3.#4s{                  
                      \relax
                      \ifmmode #1^{\rm h}\,#2^{\rm m}\,#3\fs#4
                      \else \hbox{$#1^{\rm h}\,#2^{\rm m}\,#3\fs#4$}
                      \fi
                     }
\def\dmsd#1d#2m#3.#4s{                  
                      \relax
                      #1\degr\,#2\arcmin\,#3\farcs#4
                     }
\def\mag{\relax                          
        \ifmmode ^{\rm m}
        \else $^{\rm m}$
        \fi
       }
\def\magd#1.#2{                          
              \relax
              \ifmmode #1^{\rm m}
                       \hskip-0.55em.\hskip0.22em#2
              \else \hbox{#1$^{\rm m}
                    \hskip-0.55em.\hskip0.22em$#2}
              \fi
             }

\ifoldfss
  \ifCUPmtlplainloaded \else
    \NewTextAlphabet{textbfit} {cmbxti10} {}
    \NewTextAlphabet{textbfss} {cmssbx10} {}
    \NewMathAlphabet{mathbfit} {cmbxti10} {} 
    \NewMathAlphabet{mathbfss} {cmssbx10} {} 
  \fi
  \ifAMStwofonts
    \ifCUPmtlplainloaded \else
      \NewSymbolFont{upmath} {eurm10}
      \NewSymbolFont{AMSa} {msam10}
      \NewMathSymbol{\upi}     {0}{upmath}{19}
      \NewMathSymbol{\umu}     {0}{upmath}{16}
      \NewMathSymbol{\upartial}{0}{upmath}{40}
      \NewMathSymbol{\leqslant}{3}{AMSa}{36}
      \NewMathSymbol{\geqslant}{3}{AMSa}{3E}

    \fi
  \fi
\fi 

\ifnfssone
  \newmathalphabet{\mathit}
  \addtoversion{normal}{\mathit}{cmr}{m}{it}
  \addtoversion{bold}{\mathit}{cmr}{bx}{it}
  \newmathalphabet{\mathbfit} 
  \addtoversion{normal}{\mathbfit}{cmr}{bx}{it}
  \addtoversion{bold}{\mathbfit}{cmr}{bx}{it}
  \newmathalphabet{\mathbfss} 
  \addtoversion{normal}{\mathbfss}{cmss}{bx}{n}
  \addtoversion{bold}{\mathbfss}{cmss}{bx}{n}
  \ifAMStwofonts
    \ifCUPmtlplainloaded \else
      \UseAMStwoboldmath
      \makeatletter
      \new@mathgroup\upmath@group
      \define@mathgroup\mv@normal\upmath@group{eur}{m}{n}
      \define@mathgroup\mv@bold\upmath@group{eur}{b}{n}
      \edef\UPM{\hexnumber\upmath@group}
      \new@mathgroup\amsa@group
      \define@mathgroup\mv@normal\amsa@group{msa}{m}{n}
      \define@mathgroup\mv@bold\amsa@group{msa}{m}{n}
      \edef\AMSa{\hexnumber\amsa@group}
      \makeatother
      \mathchardef\upi="0\UPM19
      \mathchardef\umu="0\UPM16
      \mathchardef\upartial="0\UPM40
      \mathchardef\leqslant="3\AMSa36
      \mathchardef\geqslant="3\AMSa3E
    \fi
  \fi
\fi 

\ifnfsstwo
  \DeclareMathAlphabet{\mathbfit}{OT1}{cmr}{bx}{it}
  \SetMathAlphabet\mathbfit{bold}{OT1}{cmr}{bx}{it}
  \DeclareMathAlphabet{\mathbfss}{OT1}{cmss}{bx}{n}
  \SetMathAlphabet\mathbfss{bold}{OT1}{cmss}{bx}{n}
  \ifAMStwofonts
    \ifCUPmtlplainloaded \else
      \DeclareSymbolFont{UPM}{U}{eur}{m}{n}
      \SetSymbolFont{UPM}{bold}{U}{eur}{b}{n}
      \DeclareSymbolFont{AMSa}{U}{msa}{m}{n}
      \DeclareMathSymbol{\upi}{0}{UPM}{"19}
      \DeclareMathSymbol{\umu}{0}{UPM}{"16}
      \DeclareMathSymbol{\upartial}{0}{UPM}{"40}
      \DeclareMathSymbol{\leqslant}{3}{AMSa}{"36}
      \DeclareMathSymbol{\geqslant}{3}{AMSa}{"3E}
    \fi
  \fi
\fi 

\ifCUPmtlplainloaded \else
  \ifAMStwofonts \else 
    \def\upi{\pi}
    \def\umu{\mu}
    \def\upartial{\partial}
  \fi
\fi

\title[\hii\ regions in M100]{Statistical properties of \hii\ regions in
the disc of M100}
\author[J. H. Knapen]{J. H. Knapen\\
Department of Physical Sciences, 
University of Hertfordshire, Hatfield,
Herts AL10 9AB, UK. E-mail
knapen@star.herts.ac.uk}

\date{Accepted January 1998;
      Received;
      in original form}

\pagerange{\pageref{firstpage}--\pageref{lastpage}}
\pubyear{1997}

\begin{document}

\maketitle

\label{firstpage}

\begin{abstract}

From a new mosaic image in the \ha\ line of the complete disc of the
spiral galaxy M100, a catalogue is composed listing 1948 individual
\hii\ regions. I give details of the data collection and reduction
procedure, and of the production of the \hii\ region catalogue. For
each \hii\ region, the catalogue gives its position relative to the
centre of the galaxy, its deprojected distance to the centre, its
radius, and its calibrated luminosity. An indication is included as to
whether the \hii\ region is located in the arms, between them, or in
the circumnuclear star-forming region. I present the results of a
statistical study of properties of the \hii\ regions. The luminosity
function of the complete ensemble of \hii\ regions shows a
characteristic shape well fitted by a power-law slope in the higher
luminosity range, and complying with literature values for galaxies
like M100. Luminosity function slopes for arm and interarm \hii\
region populations separately are found to be equal within the errors
of the fits, indicating that whereas the density wave accumulates
material into the arm regions, and may trigger star formation there,
it does not in fact change the mass distribution of the star-forming
clouds, nor the statistical properties of the \hii\ region
population. Diameter distributions and the radial number density
distribution are discussed. The latter indicates those areas where
most star formation occurs: the circumnuclear region and the spiral
arms.  The huge number of \hii\ regions allowed the construction of a
number of independent luminosity functions at different distances to
the nucleus. The slope of the luminosity function shows a marginal
decrease with increasing distance from the centre, which could
indicate a gradual change toward shallower IMF slopes with increasing
galactocentric distance, or an evolutionary effect.

\end{abstract}


\section{Introduction}

\hii\ regions in galaxies are formed when the neutral gas surrounding 
a region of massive star formation (SF) is being ionized by the
ultraviolet radiation of the young stars within it. The absorbed
radiation is emitted as a nebular emission line spectrum, with the
\ha\ line as its most powerful emission line in the optical
range. Most \hii\ regions are thought to be ionization bounded, in
which case the luminosity in e.g. the \ha\ line scales directly with
the number of ionising photons (Kennicutt 1992), although the largest
\hii\ regions, at luminosities $L>10^{38.7}$\,erg\,s$^{-1}$, are
probably density bounded (Beckman et al. 1998).  The \ha\
line is one of the most widely used lines to study massive SF in
external galaxies due to the ease with which it can be measured over
complete galactic discs.

Kennicutt, Edgar \& Hodge (1989) studied \hii\ region populations in a
sample of 30 galaxies, ranging in morphological type from Sb to
Irr. They constructed luminosity functions (LFs) and found that these
can be described by a power law function with $N(L) \propto L^{-2 \pm
0.5}$. \hii\ regions are more prevalent in late type than in early
type galaxies, and the LF is shallower in the former. Kennicutt et
al. (1989) could construct separate arm and interarm LFs for five
galaxies in their sample, and found that the interarm LF slope is
significantly different from that of the arm LF.

In a series of papers, of which the present one forms part, we have
described results from imaging spiral galaxies in the \ha\ line using
the 4.2m William Herschel Telescope (WHT). Deep images with superb
spatial resolution allowed the creation of large catalogues of \hii\
regions for individual galaxies, and a detailed study of their
statistical properties. Cepa \& Beckman (1989, 1990a) studied the
distribution of \hii\ regions in NGC~3992 and in the inner
$3\min\times4\min$ of M100. Knapen et~al. (1993a) described a detailed
study of the \hii\ regions in NGC~6814, and Rozas, Beckman \& Knapen
(1996a) and Rozas, Knapen \& Beckman (1996b) studied NGC~157,
NGC~3631, NGC~6764 and NGC~6951. This series of papers is complemented
by the study of Rand (1992) of the \hii\ regions in M51, which is
based on an image of comparably good characteristics. LFs for all
these galaxies can be fitted by power laws, complying with earlier
work. Diameter and radial distributions also show few
surprises. Significantly different slopes of arm and interarm LFs were
only found in M51. Other recent \hii\ region studies include those by
Ryder \& Dopita (1993) for nearby southern spirals, Hodge \& Miller
(1995) for local group galaxies, Tsvetanov \& Petrosian (1995) for
Seyfert galaxies, Crocker, Baugus \& Buta (1996) for ringed galaxies,
and Evans et al. (1996) and Gonz\'alez Delgado et al. (1997) for AGN
host galaxies.  \hii\ region studies have also been published for
M100: Hodge \& Kennicutt (1983) present positional data for 286 \hii\
regions from a photographic image, used by Anderson, Hodge \&
Kennicutt (1983) to analyze aspects of the galaxy's spiral structure.
Cepa \& Beckman (1990a) catalogued 456 \hii\ regions in the inner
region of the disc of M100, Arsenault, Roy \& Boulesteix (1990)
measured parameters for 127 \hii\ regions from an \ha\ image obtained
from their Fabry-P\'erot data set, and Banfi et al. (1993) catalogued
83 \hii\ regions from their \ha\ image.

In the present paper, I use a new \ha\ image of the complete disc of
M100, at uniformly high spatial resolution, to construct a catalogue
of almost 2000 individually measured \hii\ regions, and to study basic
statistical properties.

M100 (=NGC~4321) is a grand-design galaxy of type .SXS4.. (de
Vaucouleurs et al. 1991) with a moderately strong bar.  In previous
papers, we have discussed the \hi\ distribution and kinematics (Knapen
et al. 1993b) and described CO measurements (Cepa et al. 1992; Knapen
et al. 1996), which was combined with the \ha\ image described in
detail in the present paper to determine massive SF efficiencies in
arm and interarm regions (Knapen et al. 1996). The galaxy has a
circumnuclear region (CNR) of strongly enhanced SF, where a pair of
miniature spiral armlets occurs in a resonance region between the
inner and outer parts of the bar (Knapen et al. 1995a,b). I adopt a
distance to M100 of $D=16.1\pm1.3$\,Mpc (Ferrarese et al. 1996). At
this distance, 1\sec\ corresponds to $78\pm6$\,pc.

The structure of this paper is as follows. The observations and data
reduction procedures are described in Section~2. The \ha\ image of
M100 and the production of the catalogue of \hii\ regions based upon
it is discussed in Section~3. Sections~4 and 5 are dedicated to the
analysis of the catalogue, dealing with the luminosity and geometrical
distribution and how these change over the disc surface. The main
conclusions are discussed in a wider context of SF processes in
galactic discs and summarized in Section~6.

\section{Observations and Data Reduction}

The \ha\ image of M100 was obtained during two observing runs with the
4.2~m WHT on La Palma, using the TAURUS
camera in imaging mode.  Since the field of view in this setup is
limited by the filter size to around 5\min\ diameter, four fields of
the galaxy were imaged, two (eastern half of M100) during the night of
27 May 1991, and two (western half) during the night of 14 March 1992.

Narrow band redshifted \ha\ filters with width of 15\AA\ were used for
the observations, centred at $\lambda_{\rm c}=6601$\AA\ for the \ha\
line observations (redshifted using the galaxy's systemic velocity
$v_{\rm sys}=1571\kms$; Knapen et al. 1993b) and at $\lambda_{\rm
c}=$6577\AA\ and 6565\AA\ for the continuum.  Exposure times were 1200
seconds for both the on-line and the continuum image on the first
night, and $2\times900$ seconds on the second night.  An EEV CCD chip
was used during both observing runs, with a projected pixel size of
$\secd 0.{279}\!$, and a size of 1180$\times1280$ pixels.

For both observing runs, the images were first bias-subtracted and
flat-fielded in a standard way. Use of dawn and dusk flatfields gave
satisfactory results, but some interference fringes are present at low
levels in the two eastern frames. The residual variations in
background level due to this are at most of order 10 counts. Compared
with the minimum ($=3\sigma$) value of around 120 in each pixel to be
included in a catalogued \hii\ region (see below), the resulting error
in the \hii\ region flux is always smaller than $\sim10$\%, and only
approaches 10\% for the smallest and faintest regions.  After
flat-fielding, a sky background estimate was made by measuring the
levels and r.m.s.  fluctuations in areas of the images which are free
of galaxy emission.  Since in each exposure the centre of the galaxy
is in one of the corners of the chip, it was always possible to find
sufficiently large areas to determine a reliable sky value.
Subtracted sky values ranged from 750 to 1000 counts in the four (two
on- plus two off-line) eastern images, and from 150 to 500 counts in
the eight western images.  The lower sky values in the latter images
are due to shorter exposure times, and a different position of the
moon (which was setting). Note that lower sky background values result
in lower residual noise in the final \ha\ images.

The on- and off-line images were aligned to better than 0.2 pixel
using fits to positions of foreground stars. These fits were also used
to check and confirm that the seeing did not change within each set of
images that were to be combined. Cosmic rays were removed from the
eastern images by hand, where pixels affected by cosmic ray hits were
removed and replaced by the average value of a number of neighbouring
pixels.  In the case of the western exposures, where sets of two
images at the same position, and imaged through the same filter, were
available, cosmic ray hits could be removed automatically by comparing
the values of identical pixels in each set of images, and replacing
unexpectedly high pixel values by an average of their neighbour's
values.  Comparison of the manual and automatic removal methods showed
that both yield completely comparable and acceptable results.  Almost
all cosmic ray hits were removed, but a few dubious cases ($<<1$\% of
the number of unambiguously identified \hii\ regions) were marked as
such and discarded in the further analysis.

One curious linear feature which is not part of the galaxy is seen at
position RA\,$\sim\hms 12h20m29s\!\!$, $\delta\sim16\deg\ 5\min$.  At
low levels, it can be traced for some three minutes of arc.  It is
almost certainly due to a meteor which burned in the earth's
atmosphere while the exposure was taken.

After aligning and cleaning the images, the continuum image was
subtracted from the \ha$+$continuum image taken at the same
position. Following the same procedure as described by Knapen et
al. (1993a) for NGC~6814, it was found that no scaling of the
continuum images was necessary. This is because pairs of images were
taken through filters with very similar bandwidths and transmissions
and in practically the same observing conditions.

\begin{figure*}
\epsfxsize=18cm \epsfbox{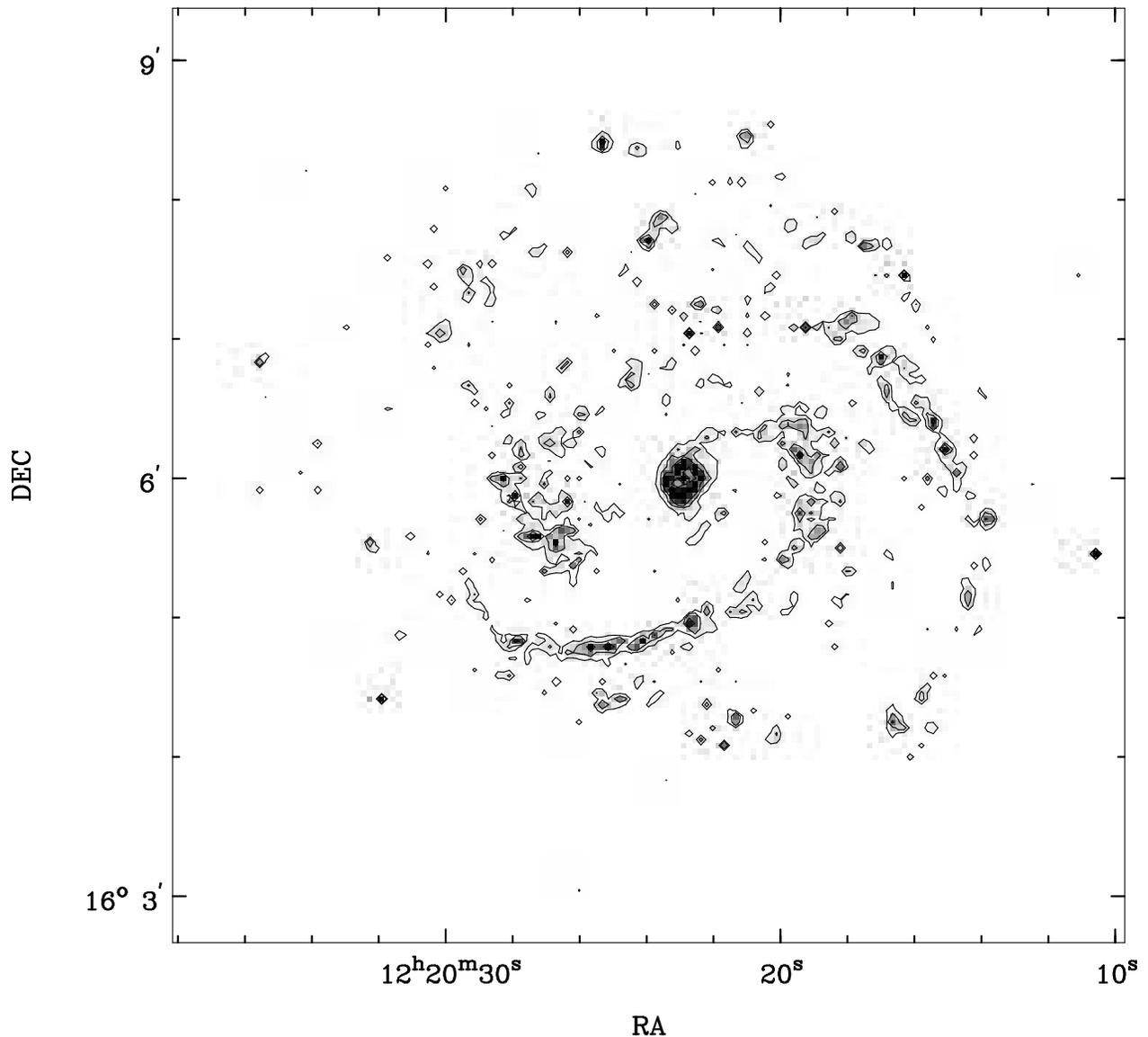} 
\caption{\ha\ continuum-subtracted image of the complete disk of M100. 
Contour levels are 0.76, 2.3, 6.8, 20.4 and
$40.8\times10^{36}$\,erg\,s$^{-1}$\,pxl$^{-1}$, grey scales range from 0.57 to
$9.5\times10^{36}$\,erg\,s$^{-1}$\,pxl$^{-1}$.\label{hamap}}
\end{figure*}

At this stage there are four sub-images of M100, with the central
region of the galaxy in another corner of the sub-image in each
case. The properties (seeing, noise, etc.) are slightly different for
each of the four, as summarized in Table~1. I used the four separate
sub-images to catalogue the \hii\ regions, and subsequently combined
the four sub-catalogues into the master catalogue, as described in
detail below.  The sub-image with the best angular resolution is the
one of the SE part of disc, and the \ha\ image of the CNR as discussed
in detail by Knapen et al. (1995a,b) was copied from that SE
sub-image. A complete image of the disc of M100 was also produced, and
is shown in Fig.~\ref{hamap}. Note that this is the image that has
been used to study massive SF efficiencies by Knapen et al. (1996) and
to study the disc morphology and relations between stars, gas and dust
by Knapen \& Beckman (1996).

\begin{table*}
  \caption{Properties of the four \ha\ sub-images obtained of the
disc of M100. Background noise, seeing and lower luminosity cutoff
for the detection of \hii\ regions are given in columns 3, 4, and 5}
\begin{tabular}{llclc}
\hline\hline
Sub-im. & Observed & Rms noise & FWHM & Cutoff\\
&& (instr. cnts.) & (\sec) & (log of erg\,s$^{-1}$)\\
\hline
\hline
NW & 14 Mar 1992 & 43 & $\secd 0.8\times\secd 0.8$ & 36.6\\
NE & 27 May 1991 & 44 & $\secd 1.0\times\secd 0.7$ & 36.7\\
SE & 27 May 1991 & 45 & $\secd 0.75\times\secd 0.65$ & 36.7\\
SW & 14 Mar 1992 & 30 & $\secd 1.0\times\secd 1.0$ & 36.5\\
\hline
\end{tabular}
\end{table*}

The four individual sub-images were combined into one mosaic as
follows. For the geometrical calibration, the sub-images were aligned
by determining the positions of the centre, residuals of field stars,
and strong, compact \hii\ regions. Photometric calibration was
available for the two nights during which the observations were taken,
but was more reliable for the 1992 run. The sub-images were scaled
relatively by determining total fluxes in certain well-defined
regions, e.g. the whole of the central region, or parts of the spiral
arms. I used several areas per pair of images considered, and found
consistent results in all cases.  The error in this relative
calibration is estimated to be $<2.5$\%. This is smaller than the
error in the photometric calibration, but in any case the dominant
errors in the absolute calibration are the uncertainties in the
distance to the galaxy and in the flux measurements of the individual
\hii\ regions.  After convolving three of the sub-images with a
Gaussian to the spatial resolution of the SW sub-image (Table~1), the
sub-images were combined into the mosaic frame, taking into account
the geometrical and calibration offsets described above. To obtain the
absolute astrometric calibration, positions of foreground stars and
the centre in the image were combined with their positions as listed
in the Hubble Space Telescope Guide Star Catalog.  The resulting image
covers the complete disc of M100, with a constant spatial resolution
of $\secd 1.0\times\secd 1.0$, and a pixel scale of $\secd0.{279}\!$.

From the photometric calibration obtained during the 1992 run, one
instrumental count in the image was found to correspond to a
luminosity of $3.78\times10^{33}$ erg/s. This is assuming the Cepheid
distance to M100 of 16.1 Mpc (Ferrarese et al. 1996). Note that the
uncertainty given for this distance (of $\pm1.3$ Mpc) corresponds to
an uncertainty of $\pm0.55\times10^{33}$ erg/s per instrumental count
in the calibration.

\section{The \hii\ region catalogue}

In the production of the \hii\ region catalogue for M100, I followed
the procedures set out before by Knapen et al. (1993a) and Rozas et
al. (1996a). Here, I briefly outline the main points, and indicate
changes with respect to the previous work. A first difference to note
is that for M100 I made separate \hii\ region catalogues for the four
sub-images, and only at a later stage combined these lists, rather
than measure all \hii\ regions from the final, combined image.

A first step is the flagging of foreground stars in the \ha\
images. This is easily done by comparison of the \ha\ and continuum
images, since stars will show comparable emission in these two, \hii\
regions, however, relatively more in \ha. This step prevents remnants
of star images in the \ha\ image to be falsely classified as \hii\
regions.  Building on our experience from previous work, I used as a
selection criterion for \hii\ regions that these should consist of at
least nine contiguous pixels, each with an intensity of at least three
times the r.m.s. noise level of the local background.  Table~1 lists
the r.m.s. noise of the background-subtracted H$\alpha$ images, and
the lower limit cutoff in the \hii\ region luminosities (determined by
multiplying the minimum number of 9 pixels, the minimum count level of
$3\sigma$, and the calibration constant), for each of the four
sub-images.  The smallest catalogued regions are 3 pixels across, and
thus have a physical diameter of some 65\,pc. Note that the detection
limit for \hii\ regions is not constant over the whole disc due to the
differences in background noise in the four sub-images. This has been
taken into account later by not fitting slopes to LFs to points below
the highest cutoff level.

After manually identifying each \hii\ region, I measured its position
in the image frame and its radius.  The flux of each \hii\ region was
determined by integrating counts within a circular aperture. Sky
subtraction errors, or errors due to local variations in the level can
influence the luminosity determinations by no more than about 10\% in
even the weakest regions (see also Sect.~2), and by a few percent in
most cases.

As discussed previously (Knapen et al. 1993a), there are possible
complications in defining the \hii\ regions and their extent due to a)
\hii\ regions partly or completely overlapping each other (as before,
each significant peak was counted as a separate \hii\ region as long
as the peaks were separated by $>3\sigma$); b) \hii\ regions not being
perfectly circular; and c) \hii\ regions having ill-defined edges as a
result of diffuse H$\alpha$ emission. Rand (1992) modelled the effects
of a) on the LF of M51, and concluded that this factor is not a
significant one in determining the true LF. I took care of b) wherever
possible by slightly adjusting the aperture over which the flux was
integrated. Finally, c) may have a slight impact on the determination
of diameters of isolated \hii\ regions, but given that edges are faint
this will not significantly influence the determination of the LF. I
thus conclude that the LF as determined from the catalogue will be a
good approximation to the ``true'' LF.

Having constructed four separate LFs with information on the raw flux
of all regions, the next step is to run a series of simple programs to
produce the final, complete, catalogue, listing absolutely calibrated
fluxes (calibration is slightly different for each sub-frame, as
described in Sect.~2), and position from the centre of the galaxy
rather than in the frame. Using values for the inclination angle of
the galaxy of 27\deg\ and for the position angle of the major axis of
153\deg\ (Knapen et al. 1993b), deprojected distances from the centre
were also calculated and listed. All the positions and distances are
listed in seconds of arc, the calibrated fluxes in
erg\,s$^{-1}$. Absolute positions of specific \hii\ regions should be
correct to within $\sim2\sec$, the accuracy of the absolute
astrometry, but relative positions among \hii\ regions should be of
much higher accuracy. I estimate the calibrated fluxes to be accurate
to $\sim10$\% on an individual basis.

All \hii\ regions in the catalogue were assigned to either arm or
interarm parts of the disc, or to the CNR. I based most of this
classification on the mask used by Knapen \& Beckman (1996). I
deviated from the mask only in the case of a string of \hii\ regions
in the bar region of M100, NW of the centre, which I classified as
``arm''. Unfortunately, these \hii\ regions are not sufficiently
numerous to warrant construction of a separate LF for bar \hii\
regions. In all, 1099 \hii\ regions were assigned to the arms, 750 to
the interarm regions, with 99 classified as forming part of the
CNR. Note that the list of 750 interarm \hii\ regions alone is several
times longer than most lists of \hii\ regions published in the
literature for complete galaxy discs so far, with only a handful of
longer lists available.

The final catalogue of \hii\ regions has 1849 entries for the disc of
M100, plus another 99 for the CNR, making a total of 1948.  The
complete list is published electronically through the CDS\footnote{The
\hii\ region catalogue is available electronically from the Centre de
Donn\'ees astronomiques de Strasbourg (CDS), on:
ftp://cdsarc.u-strasbg.fr/pub/cats/J/MNRAS/volume/first\_page .}, and
is also available in electronic form from the author.  Column 1 in the
catalogue is the number assigned to the \hii\ regions; columns 2 and 3
are the relative distances from the centre of the galaxy, in right
ascension and declination, in arcsec; column 4 is the deprojected
distance to the centre, in arcsec; column~5 the radius in arcsec; and
column~6 the luminosity of the \hii\ regions, in
$10^{36}$\,erg\,s$^{-1}$.  Finally, column 7 gives the classification
of the \hii\ regions as arm (A), interarm (I) or CNR (C).

\section{Luminosity functions}

\subsection{Total LF}

\begin{figure}
\epsfxsize=9cm \epsfbox{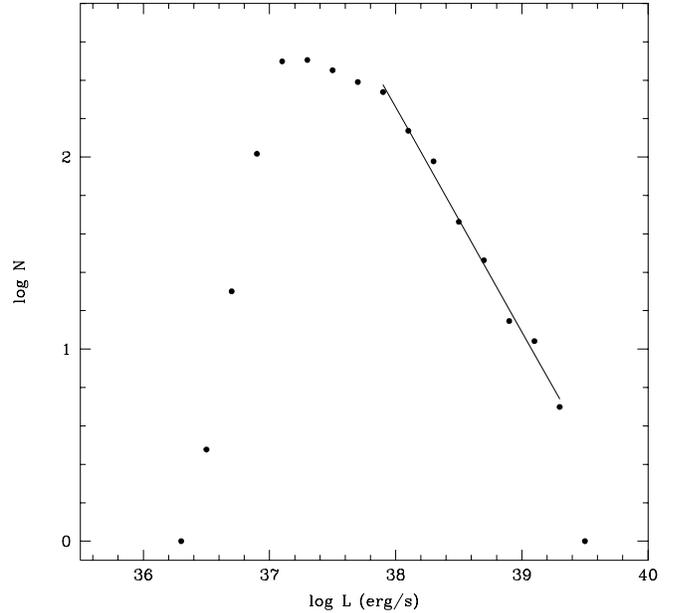}
\caption{\hii\ region LF for the disc of M100 
(excluding the CNR). Bin size is 0.2 in the log of $L$. Line indicates
best fit, and the range in $\log\,L$ over which the fit was
made.\label{disclf}}
\end{figure}

The LF of the \hii\ regions in the disc of M100 was constructed using
bins of 0.1 and 0.2 in the log of the luminosity, in order to check
possible effects of the bin size on the LF shape.  These effects were
found to be very small, and not affecting the fit of the slope to the
LF, the main parameter deduced from the LFs. I thus decided to follow
the precedent set in the literature, and show only LFs with 0.2
bin. Fig.~\ref{disclf} shows the LF for the disc of the galaxy, but
excluding the \hii\ regions in the CNR (see below). The detection
limit of individual \hii\ regions of $\log L\sim36.6$\,erg\,s$^{-1}$
is clearly visible in the LF as a drop on the low luminosity
side. This detection limit is slightly lower than that in Knapen et
al. (1993a) and similar to those reported in Rozas et al. (1996a). The
peak in the LF occurs around $\log L=37.1$\,erg\,s$^{-1}$, again
comparable to our previous studies. These similarities are of course a
result of the equal observing techniques used in all cases, but do
show the consistently high quality of the data.

I fitted a function of type $N(L)=A\,L^a\,dL$ to the high-$L$ side of
the LF, and determined a slope of the LF above $\log
L=37.9$\,erg\,s$^{-1}$ of $a=-2.17\pm0.04$. This slope is well within
the usual range of LF slopes found for galaxies of similar
morphological type (e.g. Kennicutt et al. 1989; Knapen et al. 1993a;
Rozas et al. 1996a). Note that LF slopes for M100 in the literature
range from $-1.4$ (Arsenault et al. (1990), via $-2.1$ (Banfi et
al. 1993) to $-2.7$ (Cepa \& Beckman 1990a), but are all based on data
of much lower quality.  The error on the fit to the slope is solely
the error due to the least-squares fit to the data points. The size of
the bins does not seem to be an important factor in the determination
of the slope: fitting the slope (over the same range in $\log L$) to
the LF constructed with bins of 0.1 gives $a=-2.16\pm0.04$. Including
the \hii\ regions in the CNR in the LF, predictably, lowers the slope
to $a=-1.95\pm0.04$ due to the inclusion of relatively more high-$L$
\hii\ regions. For all fits to the disc LF, the range over which the
fit was made (indicated in Fig.~\ref{disclf}) is $37.9<\log L<39.3$. A
steepening of the LF at high $L$, and/or a break or jump in the LF has
been observed in a number of spiral galaxies, in all cases near $\log
L=38.7$\,erg\,s$^{-1}$, and has been interpreted in physical terms as
a distinction between density and ionization bounded \hii\ regions
(Beckman et al. 1998). While such an effect is not immediately obvious
in the LF as shown in Fig.~2, a more detailed discussion of the
possible detection in the data set for M100 is given in Paper II
(Rozas et al., in preparation).

\subsection{Arm and interarm LFs}

\begin{figure}
\epsfxsize=9cm \epsfbox{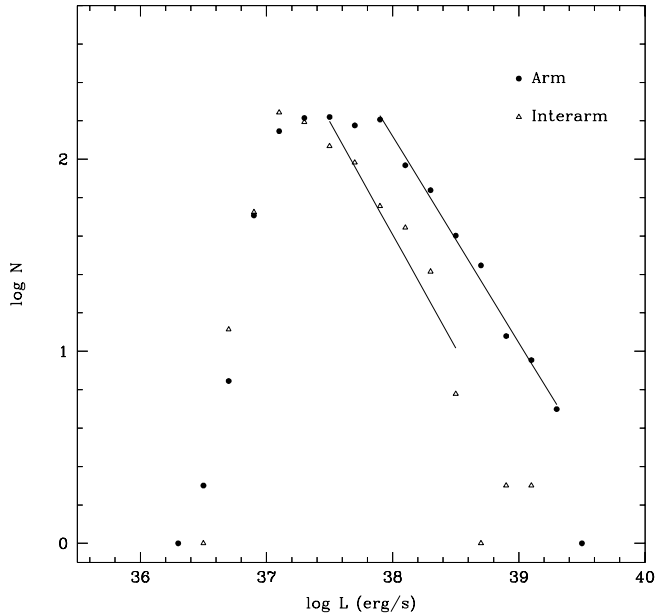}
\caption{As Fig.~\ref{disclf}, now for  arm and interarm \hii\ regions 
separately.\label{aialf}}
\end{figure}

Fig.~\ref{aialf} shows the separate LFs for the arm and interarm parts
of the disc of M100. The shapes of both the arm and interarm LF are
very similar to that of the disc LF discussed before. Fits to the
slopes of the arm and interarm \hii\ region LFs yield $a=-2.07\pm0.04$
(arm) and $a=-2.18\pm0.21$ (interarm), fitting over ranges of
$37.9<\log L<39.3$ and $37.5<\log L<38.5$, respectively. The arm and
interarm LF slopes can thus not be considered different, given the
errors in the fits. It is difficult to determine what range in $\log
L$ is best for use in the fitting, especially for the interarm LF, but
the arm and interarm slopes will not be significantly different for
any reasonable choice of range. Cepa \& Beckman (1990a) concluded from
fits to LFs, derived from their list of 456 \hii\ regions as
catalogued from an \ha\ image of the central $\mind 3.1\times\mind
3.6$ part of M100, that arm and interarm LF slopes were slightly
different. But they only used less than a quarter of the number of
\hii\ regions of the present paper, distributed over a much reduced
area of the disc. Furthermore, the errors on their fits, of about
$\pm0.3$, are large compared to the difference in slope ($-2.34$ arm,
$-3.06$ interarm).  The conclusion must be that their difference is
not significant, but does indicate the trend also found in the present
paper, that the interarm LF slope is steeper than the arm LF slope.

When compared to the arm LF, the interarm LF is slightly displaced
toward the lower left of the diagram, i.e., toward lower numbers and
lower luminosities.  However, the fact that the arm and interarm LF
slopes are equal is a strong indication that the \hii\ populations in
and between the arms are {\it not} different (see also Section~6).

\subsection{Circumnuclear region LF}

\begin{figure}
\epsfxsize=9cm \epsfbox{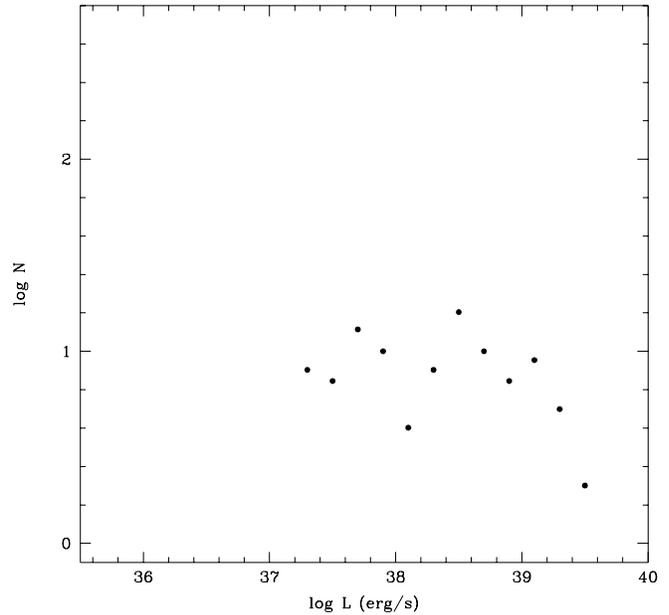}
\caption{As Fig.~\ref{disclf}, now  for the \hii\ regions in the CNR 
of M100 (inner $\sim25\sec$).\label{cnrlf}}
\end{figure}

The CNR in M100 accounts for about 16\% of the total \ha\ line flux of
the complete galaxy (Knapen et al. 1995a). The \ha\ emission is
organised into a pair of tightly wound spiral armlets (Knapen et
al. 1992, 1995a,b). When attempting to catalogue the individual \hii\
regions here, this large flux concentrated in a relatively small area
translates into a crowding problem. On the one hand, the area taken up
by the many luminous \hii\ regions precludes detection of the less
luminous \hii\ regions, whereas on the other hand, a luminous \hii\
region may in fact be a group of smaller \hii\ regions which is not
resolved. These problems were also encountered in especially arm
regions of spiral discs, but in the CNR they are much more
prominent. I did catalogue the \hii\ regions in the CNR however, and
list 98 of them in the (electronically) published catalogue.

The resulting CNR \hii\ region LF is shown in Fig.~\ref{cnrlf}. As
expected, it shows a prominent \hii\ region population at high
luminosities, whereas the low end of the LF is underpopulated to a
large extent when compared with e.g. the disc LF. It is possible that
smaller \hii\ regions are indeed less numerous in the CNR, when they
would be swept up into larger \hii\ regions, but given the
observational difficulties described above this cannot be proven here.

\subsection{LFs as a function of radius}

\begin{figure*}
\epsfxsize=16cm \epsfbox{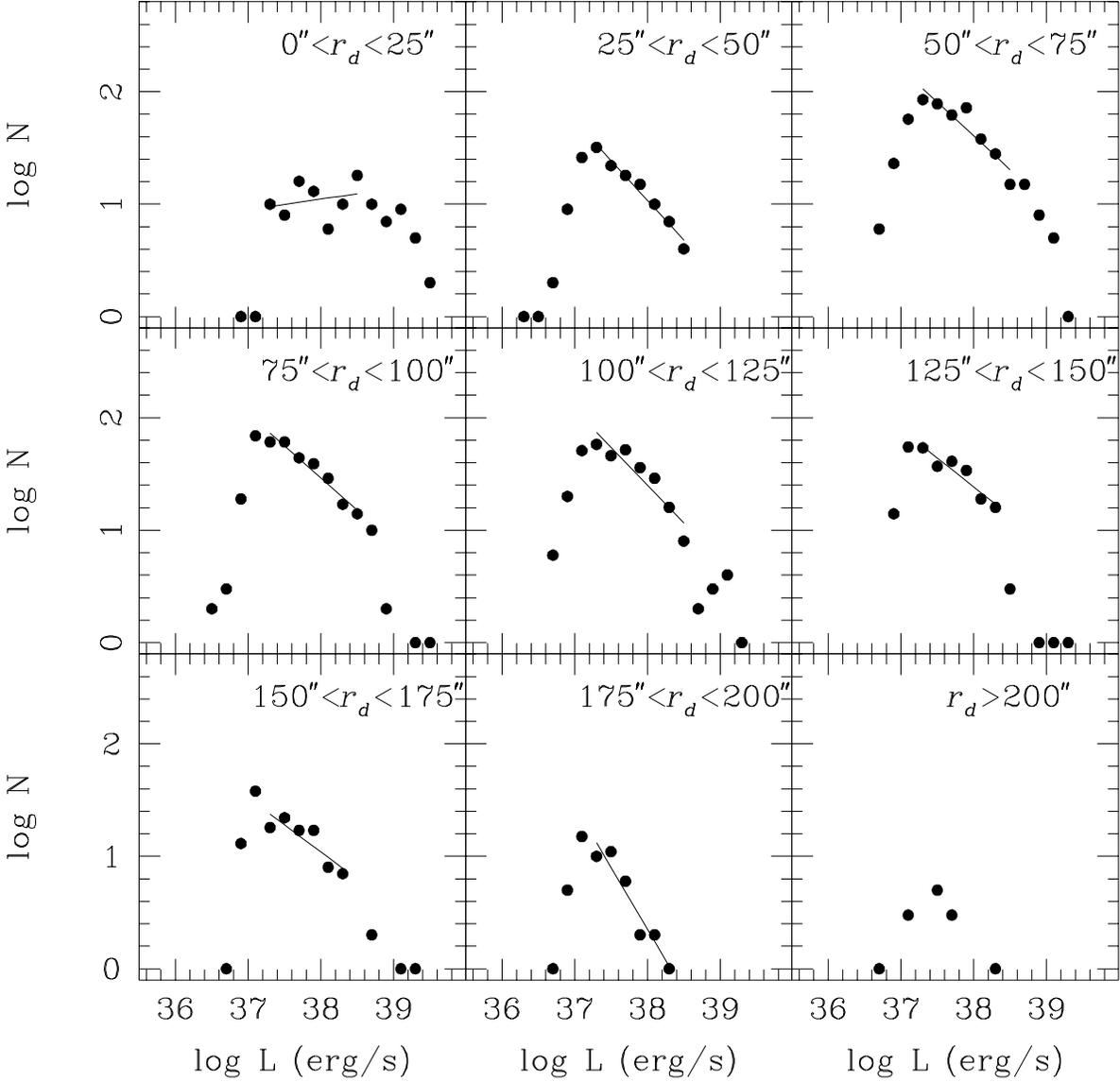}
\caption{As Fig.~\ref{disclf}, but for  \hii\ regions within ranges 
of increasing deprojected distance from the centre (bins are 25\sec\
wide). The appropriate distance range for each LF is indicated in the
upper right corner of each panel.\label{lfrad}}
\end{figure*}

The large number of \hii\ regions in the catalogue of M100 allowed me
to perform a novel type of analysis: to study the detailed radial
dependence of the LFs in the disc. There are a number of reports in
the literature on LFs in the ``inner'' and ``outer'' parts of discs
(e.g. Kennicutt et al. 1989; Rand 1992; Knapen et al. 1993a; Banfi et
al. 1993), where the division between these sections is usually placed
around $0.5\,R_{25}$. Rand (1992) reported different slopes for inner
and outer parts of the disc of M51, but in the other cases {\it no}
significant differences were found.

\begin{table}
\caption{LF slopes as function of radius. $L$ is in erg\,s$^{-1}$.}
\begin{tabular}{llrcc}
\hline
\hline
Radial bin & Central $r_{\rm d}$ & No. & Fitted range & LF slope $a$\\
(\sec) & (\sec) & & ($\log L$) & \\ 
\hline 
\hline 
0--25 & 12.5 & 116 & 37.3--38.5 & $-0.91\pm0.17$\\ 
25--50 & 37.5 & 147 & 37.3--38.5 & $-1.71\pm0.05$\\ 
50--75 & 62.5 & 493 & 37.3--38.5 & $-1.60\pm0.11$\\
75--100 & 87.5 & 373 & 37.3--38.5 & $-1.57\pm0.06$\\  
100--125 & 112.5 & 332 & 37.3--38.5 & $-1.67\pm0.12$\\ 
125--150 & 137.5 & 276 & 37.3--38.3 & $-1.51\pm0.09$\\ 
150--175 & 162.5 & 145 & 37.3--38.3 & $-1.48\pm0.14$\\ 
175--200 & 187.5 & 53 & 37.3--38.5 & $-2.10\pm0.16$\\
200--225 & 212.5 & 13 & -- & --\\ 
\hline
\end{tabular}
\end{table}

Fig.~\ref{lfrad} shows the results of the more objective and detailed
analysis made possible by the number of almost 2000 \hii\ regions in
the M100 catalogue. I constructed LFs, exactly in the same way as done
before for e.g. the whole disc, for nine different ranges or bins of
deprojected distance from the centre $r_{\rm d}$, with width 25\sec, and
starting at $r_{\rm d}=0\sec$ (central position). The last bin includes all
\hii\ regions lying further from the centre than 200\sec. Total
numbers of \hii\ regions included in each radial bin are decent in all
cases except the last two. Table~2 lists the number of \hii\ regions
and the fitted LF slope and error for each radial bin, as well as the
range in $L$ used for the fit.

I fitted slopes to the LFs in each radial bin. As before, the choice
of what range in $\log L$ to include in the fit is somewhat subjective
and can directly influence the results, so I chose to use exactly the
same range in all cases (with the exception of radial bins
$125\sec<r_{\rm d}<150\sec$ and $175\sec<r_{\rm d}<200\sec$, where the
inclusion of the last point leads to an unrealistic fit to the data
set, and for bin $150\sec<r_{\rm d}<175\sec$ where there is no data
point for $\log L=38.5$).

\begin{figure}
\epsfxsize=9cm \epsfbox{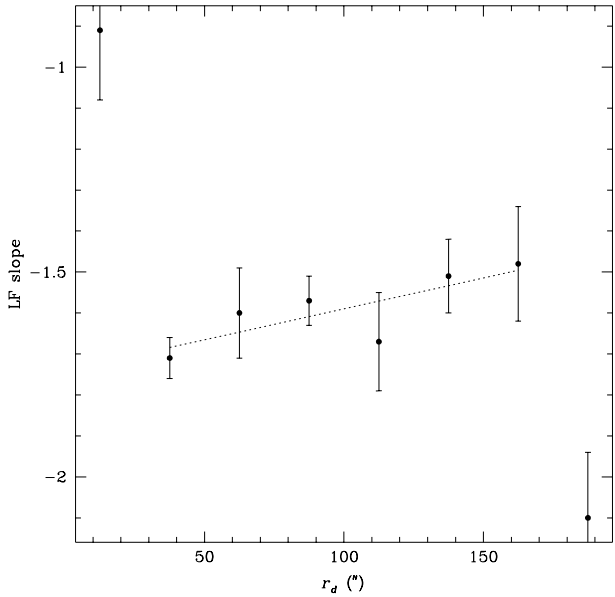}
\caption{Values of the LF slopes  fitted to the LFs shown in 
Fig.~\ref{lfrad}, plotted as a function of deprojected distance from
the centre (the central distance value of each
bin was used). Dashed line indicates best fit to the data points, and
range in distance used for the fit.\label{sloperad}}
\end{figure}

The results are listed in Table~2 and plotted as a function of $r_{\rm
d}$ in Fig.~\ref{sloperad}. Fig.~\ref{sloperad} also shows the fit to
the radial behaviour of the LF slopes.  The LF slopes become shallower
with increasing $r_{\rm d}$, although this is hardly significant
taking the errors on the fits into account. Two discrepant values of
the fitted slope occur for the first radial bin, $r_{\rm d}<25\sec$,
and for bin no.~8, $175\sec<r_{\rm d}<200\sec$. In the first bin, the
LF is more or less the LF as derived before for the CNR, which is
basically flat up to very high $L$.  The small number of \hii\ regions
in bin no. 8 is probably the cause of the deviant LF slope in that
bin.

\section{Diameter and radial distributions}

\subsection{Diameter distribution}

\begin{figure}
\epsfxsize=9cm \epsfbox{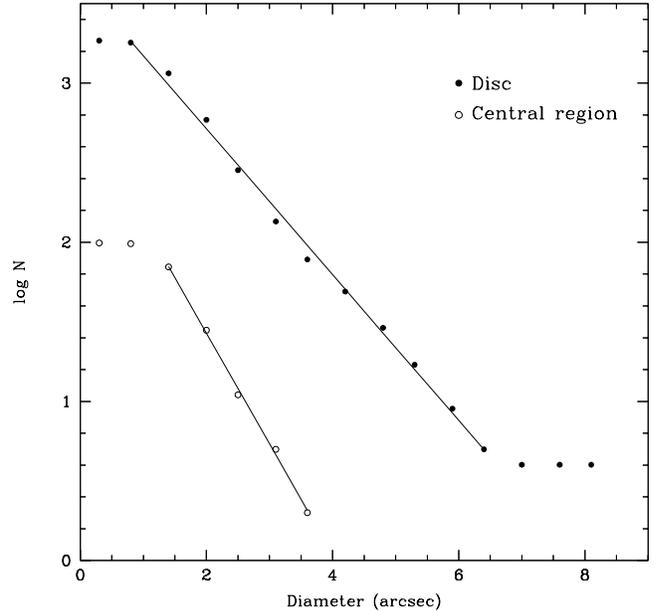}
\caption{Integral diameter distribution for the \hii\ regions in 
the disc of M100 (but excluding those in the CNR; filled dots), and in
the CNR (open symbols). Lines indicate exponentials fitted to the data
points, where the extent of the plotted line indicates the data points
used in the fit.\label{diams}}
\end{figure}

The integral diameter distribution of \hii\ regions in spiral galaxies usually follows an exponential law:

$$N(>D)=N_0\,e^{-D/D_0}$$

with $N(>D)$ the total number of \hii\ regions with diameters larger
than $D$ (van den Bergh 1981; Hodge 1987; Ye 1992). The slope of the
diameter distribution is correlated with the luminosity of the galaxy
(Hodge 1987; see also Rozas et al. 1996b), and the slopes tend to be
steeper for \hii\ regions located between the spiral arms than for
those within them in the few cases where this has been explicitly
studied (Hodge 1987; Knapen et al. 1993a).

\begin{figure}
\epsfxsize=9cm \epsfbox{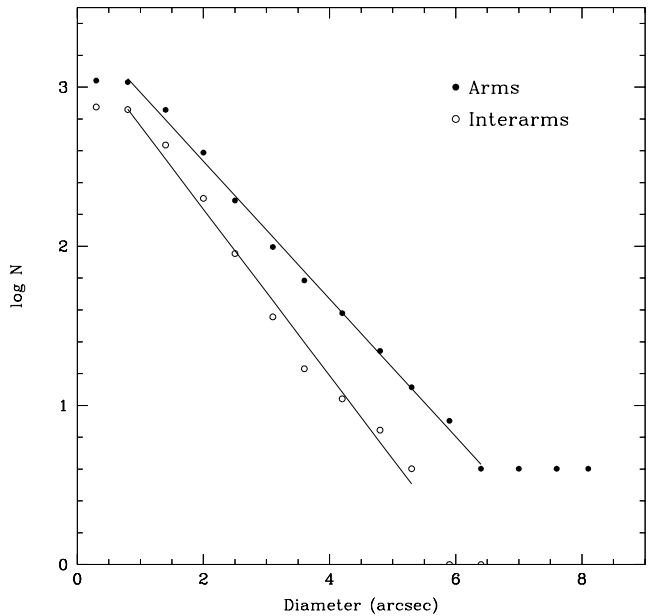}
\caption{As Fig.~\ref{diams}, now for arm (filled dots) and interarm 
(open symbols) \hii\ region populations separately.\label{aiadiams}}
\end{figure}

The integral diameter distribution of the \hii\ regions in the disc of
M100 is shown in Fig.~\ref{diams} (full dots).  Apart from the first
point, and the last three points in the Figure, the data are fitted
very well with an exponential of the type described above.  As usual,
the fit to the data points is indicated in the Figure.  I also show
integral diameter distributions for the \hii\ regions in the CNR (open
circles in Fig.~\ref{diams}), and in the arm and interarm areas of the
disc separately (Fig.~\ref{aiadiams}).  In all cases the data points
can be well fitted with an exponential of the type described above.
The fitted values to the slopes are $D_0=74\pm2$pc and
$N_0=(4.27\pm0.37)\times10^3$ for the whole disc, $48\pm2$pc and
$(0.66\pm0.08)\times10^3$ for the CNR \hii\ regions only, and
$78\pm2$pc and $(2.52\pm0.19)\times10^3$; and $65\pm3$pc and
$(1.90\pm0.34)\times10^3$ for the arm and interarm \hii\ regions,
respectively.

The smaller size scale in the interarm region could be due to
evolutionary effects, the interarm \hii\ regions being more evolved
(Oey \& Clarke 1998) and smaller since their Stromgren sphere will
have shrunk. More likely, however, the smaller interarm value for
$D_0$ is due to the fact that in the arms \hii\ regions overlap more
often, and determining the diameters of the smaller \hii\ regions will
be more difficult.

The disc value derived here is significantly smaller than values
derived by other authors: after correcting to the distance to M100 as
used in the present paper, Arsenault et al. (1990) find $D_0=112$~pc,
Cepa \& Beckman (1990a) $D_0=128$~pc, and Banfi et al. (1993)
$D_0=165$~pc (no erros given). Especially in the case of Banfi et
al. (1993), who catalogue only 83 \hii\ regions for M100, the larger
values found in the literature can be explained by a smaller number of
\hii\ regions used as input for the diameter determination. This is
usually a result of a combination of lower resolution and lower
sensitivity in the \ha\ images, leading to overestimates of the
diameters of \hii\ region complexes which would have been resolved
using better imaging. The value of $D_0=74\pm2$~pc places M100 below
the relation found by Hodge (1987) between the absolute magnitude
$M_{\rm B}$ and the \hii\ region size scale $D_0$ for spiral
galaxies. However, the effects on systematic and instrumental
influences on the individual data points in the $M_{\rm B}$ vs. $D_0$
relation, as outlined above, remain to be investigated.

The diameter distribution for the CNR shows a steeper slope than any
of the others, and the interarm slope is significantly steeper than
the arm slope, following the trend set in earlier work (Hodge 1987;
Knapen et al. 1993a). However, especially in the CNR the diameter
determination may be less reliable due to crowding, making it harder
to determine the true extent of the \hii\ regions, and thus
underestimating the diameters of many of them. On the other hand, one
would expect the same effect to have more impact on the arm than on
the interarm \hii\ regions, so the shallower slope for the arm \hii\
regions may in fact be a lower limit to the true slope. Finally, the
fact that the LFs are very similar for the arm and interarm \hii\
region populations means that interarm \hii\ regions will have, on
average, higher surface brightnesses than arm \hii\ regions (see also
Knapen et al. 1993a).

\subsection{Radial distribution}

\begin{figure}
\epsfxsize=9cm \epsfbox{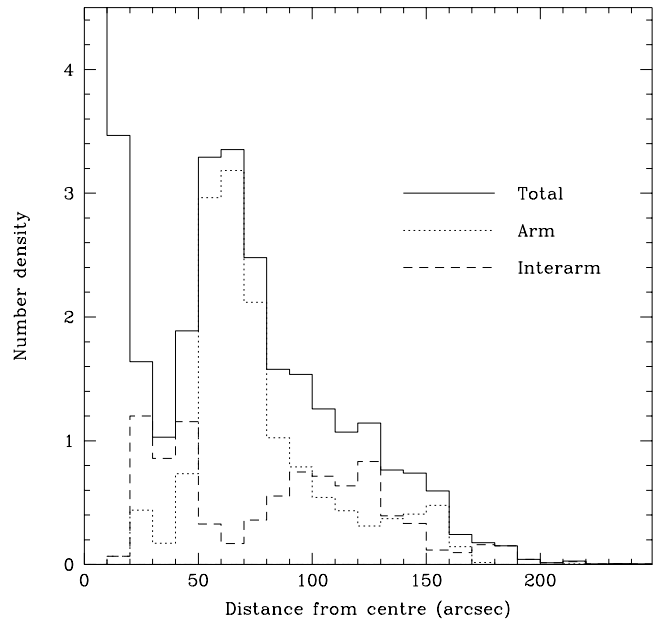}
\caption{Number density distribution for all \hii\ regions catalogued 
in M100 (including the CNR; drawn line), and for arm (dotted line) and
interarm (dashed line) \hii\ region populations separately. The number
density in the innermost bin is 9.8.\label{numdens}}
\end{figure}

Fig.~\ref{numdens} shows the number density of \hii\ regions in M100
as a function of deprojected distance from the centre. The number
density is defined as the number of \hii\ regions in annular bins of
10\sec\ width, normalized by the area in which they are found. Number
densities are shown for all \hii\ regions in the catalogue, and for
arm and interarm \hii\ regions separately. At the innermost data point
the number density is in fact 9.8, a high value caused by the large
number of \hii\ regions in the CNR. The distribution is characterised
by a number of components that can be easily recognised
morphologically, e.g. on the grey-scale image of Fig.~\ref{hamap},
namely the strongly star-forming CNR ($0\sec<r_{\rm d}<20\sec$), the
bar which is mostly devoid of massive SF ($20\sec<r_{\rm d}<50\sec$),
the star-forming spiral arms ($50\sec<r_{\rm d}<160\sec$), and the
outer disc, with only sporadically occurring \hii\ regions
($160\sec<r_{\rm d}<240\sec$). Note that these components were also
identified by Knapen \& Beckman (1996) from a radial \ha\
profile. Within the spiral arm region, the highest number density of
\hii\ regions occurs in a relatively small area with radii between
some 50\sec\ and 80\sec. This is the region where the large
star-forming complexes are located which can be seen in
e.g. Fig.~\ref{hamap} near the ends of the bar, and especially in the
spiral arm part south of the nuclear region, connecting with the
western end of the bar.  The separate arm and interarm \hii\ region
distribution shows less interarm \hii\ regions in the crowded areas in
the arms ($50\sec<r_{\rm d}<80\sec$), and more interarm than arm \hii\
regions in the bar region and at specific radii further out in the
disc (e.g. $100\sec<r_{\rm d}<130\sec$).

In general, the number density distribution is similar to the radial
profile as shown by Knapen \& Beckman (1996; their Fig.~4), where the
radial \ha\ profile is plotted logarithmically. Note that the
outermost \hii\ regions occur at distances $r_{\rm d}>D_{25}/2$ from
the centre, the outermost one lies near $r_{\rm d}\sim245\sec$.

\section{Discussion and summary}

In this paper, I present a detailed study of a new, high-quality \ha\
continuum-subtracted image of the grand-design spiral galaxy
M100. From the image, I have catalogued a total of 1948 individual
\hii\ regions, and tabulated basic properties for each \hii\ region:
position with respect to the centre, radius, and integrated flux. A
substantial part of the paper is devoted to deriving LFs for the
complete sample of \hii\ regions, and sub-samples of the whole set,
defined on the basis of the location of the \hii\ regions in the
galaxy. Following previous work, separate LFs for arm, interarm and
CNR \hii\ regions were derived, but the huge number of individual
\hii\ regions available also made it possible, for the first time, to
derive a set of nine LFs for \hii\ regions at increasing distance from the
centre.

The total LF (Fig.~\ref{disclf}) shows the by now well-known shape and
characteristics (see e.g. Kennicutt et al. 1989; Rand 1992; Knapen et
al. 1993a; and Rozas et al. 1996a for LFs of other galaxies). A sharp
cut-off at low luminosities indicates our detection limit, while the
slope at luminosities larger than the completeness limit can be well
fitted with a power law, giving an LF slope of $a=-2.17\pm0.04$, well
within the limits established in the previous works for a galaxy of
this morphological type. I fail to detect a significant difference
between the LF slopes for the arm and interarm populations of \hii\
regions. As discussed before, this does not contradict published
results by Cepa \& Beckman (1990a) who studied the central part of
M100.

This  adds yet another galaxy to the list of those that do not
show significantly different arm vs. interarm \hii\ region LF
slopes. Rand (1992) did find significantly different slopes in M51, a
galaxy with a relatively small number of interarm \hii\ regions and
strong density waves in its disc, as did Kennicutt et al. (1989) for a
number of their galaxies, although the latter result was based on
observational data of considerably lower quality than those used in
later work. Kennicutt et al. (1989) present a combined result for five
galaxies, which shows slightly steeper slopes for interarm \hii\
regions (no fits to the slopes or estimates of errors are given so the
significance of the result can not be easily compared with those for
other galaxies). Rand (1992) interpreted the significantly different
arm-interarm LF slopes in terms of a different molecular cloud mass
spectrum, but Oey \& Clarke (1998) explain such differences in general
by evolutionary effects and the maximum number of ionizing stars per
cluster.  Knapen et al. (1993a) and Rozas et al. (1996a) did not find
significant arm-interarm LF slope differences in the 5 galaxies they
studied. 
 
In most cases where interarm and arm LF slopes have been considered,
the interarm LF slopes are marginally steeper than arm LF slopes,
although not significantly so. But for two of the four galaxies
studied by Rozas et al. (1996a) the arm LF slope is steeper, so even
{\it if} the trend noticed exists, it is not unique. The only way
forward is to observe more galaxies, but significantly different LF
slopes in arm and interarm environments can be ruled out based on the
existing data, with the exception of one galaxy (M51).

This implies that the observed occurrence of many large \hii\ regions
in the spiral arms does not indicate a preference for the larger \hii\
regions to form there, but merely a statistical effect: there are more
\hii\ regions in the arms, so there will also be more large ones (e.g.
Elmegreen 1993).  The spiral density waves which must be present in
grand-design galaxies thus only re-organize the material from which
the stars form, and the SF, into the arms.  This agrees with the
observed lack of correlation between the SF rates per unit area and
arm class (Elmegreen \& Elmegreen 1986)\footnote{The SF rate per unit
area does vary strongly with morphological type, see Kennicutt \& Kent
(1983) or the review by Kennicutt (1992), and the references
therein.}.  But equal arm and interarm \hii\ region LF slopes seem
harder to unite with the observed higher massive SF efficiencies
within the arms as compared to outside the arms (Cepa \& Beckman
1990b; Knapen et al.  1992, 1996), which are prima facie evidence for
triggering of the massive SF within the arms, presumably by the
density wave.  The overall conclusion must be that {\it whereas the
density wave organizes the material into the arms}, and enhances
(triggers) the massive SF, {\it it does not in fact change the mass
distribution of the star-forming molecular clouds}, and thus of the
statistical properties of the \hii\ region population.  It just allows
more of basically the same \hii\ regions to start emitting in the
arms.

There is a slight change in the LF slopes with increasing radius
(Fig.~\ref{sloperad}), with slopes decreasing (becoming shallower)
with increasing radius. Note that Rand (1992) found a steeper slope in
the outer part of the disc of M51 than in the inner part, a result
which goes in the opposite direction as the trend found here for M100.
The result for M100 would indicate the presence of relatively more
large \hii\ regions at larger distances from the centre, or, following
the results of the theoretical treatment by Rozas (1996), a shallower
IMF further out in the disc. Alternatively, evolutionary effects in
the \hii\ region population, as described by e.g. Oey \& Clarke
(1998), could be responsible for the observed effect.  This is an
interesting point which should be followed up by more extensive study,
preferably through spectroscopic observations of decent numbers of
\hii\ regions at varying distances from the centre.

{\it Acknowledgements} 

I thank J.E.  Beckman for his continuing support of and interest in this
work, and M.S.  Oey and C.H.  Heller for helpful discussion.  N. 
Arnth-Jensen wrote the first version of most of the Fortran programmes I
used.  This paper is based on observations obtained at the William
Herschel Telescope, operated on the island of La Palma by the Royal
Greenwich Observatory in the Spanish Observatorio del Roque de los
Muchachos of the Instituto de Astrof\'\i sica de Canarias.  Financial
support from the British Council and the Spanish Acciones Integradas
Programme, and from the Spanish DGICYT, Grant No.  PB94-1107, is
acknowledged.

\bsp

\label{lastpage}

\end{document}